\documentstyle[12pt]{article}                               

\def\be{\begin{equation}}
\def\en{\end{equation}}

\begin{document}
\begin{titlepage}
\baselineskip = 25pt
\begin{center} 

{\bf On the sources of the late integrated Sachs-Wolfe effect}
 
\vspace{.5 cm}
 
M\`arius J. Fullana$^{1,2}$ and Diego S\'aez$^{1}$ \\   
\small
$^{1}$Departamento de Astronom\'{\i}a y 
Astrof\'{\i}sica. Universidad de Valencia.\\
46100 Burjassot, Valencia, Spain\\
$^{2}$Departament de Matem\`atica Aplicada. Universitat
Polit\`ecnica de Val\`encia.\\
46071 Val\`encia.\\
\footnotesize
e-mail: mfullana@mat.upv.es; diego.saez@uv.es\\
\end{center}

\vspace {2. cm}
\normalsize
\begin{abstract}

In some scenarios,
the peculiar gravitational potential of
linear and mildly nonlinear structures depends on time and, as 
a result of this dependence, a  
late integrated Sachs-Wolfe effect appears. 
Here, an appropriate formalism is used which 
allows us to improve on the analysis of the spatial scales 
and locations of the
main cosmological inhomogeneities producing this effect.
The study is performed in the framework of the
currently preferred flat model with cosmological constant,
and it is also developed in an open model for comparisons.
Results from this analysis are used to discuss 
the contribution of Great Attractor-like objects,
voids, and other
structures to the CMB anisotropy.

\end{abstract}

{\em Subject headings:} cosmic microwave background---cosmology:theory---
large-scale structure of the universe---methods:numerical


\end{titlepage}

\section{INTRODUCTION}       

In some scenarios where linear and mildly nonlinear structures create a
time varying gravitational potential, the photons of the Cosmic
Microwave Background (CMB) undergo a late Integrated Sachs-Wolfe
(ISW) effect. In the absence of any cosmological constant, the partial 
time derivative of the gravitational potential tends to zero as the 
universe approaches a flat one and, consequently, the ISW effect 
tends also to zero. This paper is devoted to the study of some aspects 
poorly known of
the ISW effect: our goal is a detailed analysis of the locations and scales
of the subhorizon structures 
contributing to this effect.
We are particularly 
interested in the scales corresponding 
to observable objects as voids, the Great wall, et cettera and,
by this reason, we will only consider scales smaller than 
the horizon.
We choose an adequate formalism to deal with this analysis.

Two scenarios are considered: a flat universe with 
cold dark matter (CDM) and cosmological constant
and an open universe with CDM, 
they are hereafter referred to as scenarios (or models)
I and II, respectively. In both cases the spectrum has been first normalized
by the condition $\sigma_{8}=1$, to consider other normalizations 
(other $\sigma_{8}$ values) when necessary.
In case I, 
the spectrum corresponds to cold dark matter (CDM) with
$\Omega_{d}=0.25$, $\Omega_{b}=0.05$,
$\Omega_{\lambda}=0.7, $ $h=0.65$ and $n=1$,
where                            
$\Omega_{b}$, $\Omega_{d}$, and $\Omega_{\lambda}$
are the density parameters corresponding 
to baryonic matter, dark matter, and the cosmological constant,
respectively, 
$h$ is the reduced Hubble constant 
($h=H_{0}/100$, $H_{0}$ being the Hubble constant in units of 
$Km/s.Mpc$), and
$n$ is the spectral index
of the primordial scalar energy density fluctuations. 
The scenario II
involves CDM and the relevant parameters are:
$\Omega_{d}=0.25$, $\Omega_{b}=0.05$, $h=0.65$ and $n=1$.  
Model I is currently preferred according to recent observations
of far Ia supernovae and the CMB spectrum (location of the Doppler
peak), while model II
can account for the abundances of rich clusters and Einstein's rings
and, here, it is mainly used for comparisons.
As it is well known, the normalization $\sigma_{8}=1$
does not lead to a good normalization of the CMB angular power
spectrum in most cases; in other words, when the $C_{\ell}$
coefficients are calculated (for $\sigma_{8}=1$), 
the resulting values do not fit well with the 
values observed by COBE, TENERIFE and other experiments. In each scenario,
appropriate fits to the observed CMB spectrum correspond
to $\sigma_{8}$ values
which are, in general, different from unity. The so-called
bias parameter is $b= 1/ \sigma_{8}$.

Since our attention is focused on subhorizon scales, we will
estimate the late ISW anisotropy in the $\ell$--interval $(10, 40)$.
For $\ell < 10$, super-horizon scales would be important
(see Kamionkowski and Spergel, 1994) and, then, the spatial curvature
could be only neglected in model I; furthermore,  
the cosmic variance would lead to important 
uncertainties ($\Delta C_{\ell} / C_{\ell}$ is proportional 
to $[2/(2 \ell + 1)]^{1/2}$, see Knox, 1995)
and the Sachs-Wolfe effect would be very important.
For $\ell > 40$, the Doppler effect starts its domination
hidding other effects as the ISW one.
An appropriate linear approach is used in next sections 
to estimate the $C_{\ell}$ 
coefficients for $10 \leq \ell \leq 40$. The method used to do this
estimation should facilitate the separation of the ISW effect 
from other contributions
to the angular power spectrum and, moreover, this method should give
information about the sizes and locations of the main subhorizon 
structures 
contributing to the late ISW effect. 
The numerical integration of the Boltzmann equation
or the computational strategy of Hu and Sugiyama (1994) 
could be used to perform
the analysis of this paper; nevertheless, 
another appropriate approach --based on a certain approximation 
to the sources-- is described and used in next sections.

In previous papers, 
it was claimed that 
some Great Attractor-Like (GAL) objects located between
redshifts $2$ and $30$ in open enough universes (without cosmological
constant)
could account for 
an important part of the
Integrated Sachs-Wolfe (ISW) effect. 
Arguments in those papers were based on the 
Tolman-Bondi (TB) solution of Einstein's equations, which
was used to estimate 
both the anisotropy produced by a single GAL structure 
(Arnau, Fullana \& S\'aez, 1994; S\'aez, Arnau \& Fullana, 1995)
and the abundance of these structures (S\'aez \& Fullana 1999). 
Unfortunately,
our TB simulations
have some features, as the spherical symmetry
and a particular form of compensation, which
could affect abundance and anisotropy estimations.
By this reason,
the mentioned claim should be discussed using a general formalism 
(not TB solution). It is done in Section 4 as a subsidiary
application (in model II) 
of the formalism described along the paper.
 
In section 2, the method used to compute the 
angular power spectrum inside the $\ell$ interval [10, 40] is
described. Results are presented in Section 3 and, Section 4
is a general discussion and a summary of conclusions.
Finally, some words about notation:
whatever quantity "$A$" may be, $A_{_{L}}$ and $A_{0}$ stand for
the $A$ values on 
the last scattering surface and at present time,
respectively. Simbols $x^{i}$,
$\phi$, $\vec {v}$, $\vec {n}$, $\rho_{_{B}}$, $\delta$, $a$, $t$, $G$,
stand for the comoving coordinates, the peculiar gravitational potential,  
the peculiar velocity, the unit vector in the observation direction,
the background mass density, the density contrast, the scale factor,
the cosmological time, and the gravitational constant, respectively. 
Units are chosen in such a way that the speed of light is $C=1$.
Quantities
$\Omega_{0}$ and $\Omega_m$ are defined as follows:
$\Omega_{0} = \Omega_{b} + \Omega_{d} + \Omega_{\lambda}$ and
$\Omega_{m} = \Omega_{b} + \Omega_{d}$. The 
comoving wavenumber is $k_{c}$, while $k$ is the physical one.

\section{$C_{\ell}$ ESTIMATIONS IN THE $\ell $ INTERVAL [10,40]}

We are interested in the ISW effect produced by structures much smaller
than the horizon scale and, consequently, the region of 
the hypersurfaces
$t=constant$ where the inhomogeneities interact 
with the CMB can be considered as flat; 
namely, the
spatial curvature can be neglected in the open model II. 
This means that, even in the open case, 
the spatial part of the functions defining the linear structures
under consideration
can be expanded in plane waves. It is not necessary the use of
the complicated solutions of the Helmholtz equation, which
should be used in open backgrounds to do an exact
and rigorous treatment of structure evolution. 
Of course, in model II, the {\em time evolution} is studied 
taking into account the
existence of a {\em space-time} curvature distinguishing the open 
universe from the flat one. 

The potential approximation is used in our
estimates. The basic  equations
(Mart\'{\i}nez-Gonz\'alez et al. 1994, 
Sanz et al. 1996) are:
\be
\frac {\Delta T} {T} = 
\frac {1}{3} \phi_{_{L}} + \vec {n} \cdot \vec {v}_{_{L}} +
2 \int_{t_{e}}^{t_{o}} dt \frac {\partial \phi}
{\partial t}   \ , 
\label{efun}
\en
and
\be
\Delta \phi = 4 \pi G \delta a^{2} \rho_{_{B}}  \ , 
\label{lapla}
\en
where $\frac {\Delta T} {T}$
is the relative temperature variation 
--with respect to the background temperature-- along the  
direction $\vec {n}$, and 
the integral is to be computed from emission ($e$) to
observation ($o$) along the background null geodesics.
Initially, this approach was designed to study the flat case
(scenario I); nevertheless, 
the potential approximation can be also used in the
open case (scenario II) provided that we are concerned with
structures smaller than the horizon scale (Sanz et al. 1996).
The first, second, and third terms of this equation give
the Sachs-Wolfe, the Doppler and the ISW
anisotropies, respectively. Hereafter, we write
$A^{^{S}}$, $A^{^{D}}$, or $A^{^{I}}$ to indicate that the 
quantity $A$ has
been estimated using only the first, the second or the
third term, respectively. 

In the linear pressureless approach,
the density contrast evolves as follows:
\be
\delta (x^{i}, t)= \frac {D_{1} (a)}{D_{1} (a_{0})} 
\delta (x^{i}, t_{0})  \ , 
\label{l1}
\en
where  
$D_{1}(a)$ describes the evolution of the growing mode 
of the density contrast  
(Peebles, 1980). 
The form of the function $D_{1}(a)$ is different in models I
and II. Hereafter, $D_{1q}(a)$ stands for 
the functions $D_{1}(a)$ corresponding to our two scenarios. The same
notation based on the subscript $q$ is also used for other quantities.
This subscript $q$ is only used along the text
to distiguish the model I (q=I) from the model II (q=II). 
Functions $D_{1q}(a)$ and all the quantities having the 
subscript $q$ are written in Appendix A for 
q=I and q=II.

The peculiar gravitational potential can be written as
follows 
\be
\phi (x^{i}, t) = \Phi (x^{i})  \frac {D_{1}(a)}{a}  \ , 
\label{l4}       
\en
where function $\Phi(x^{i})$ satisfies the equation
$\Delta \Phi = B_{q} \delta (x^{i}, t_{0})$. 

Let us focus our attention on
the angular power spectrum of the CMB; namely, on the coefficients
\be
C_{\ell} = \frac {1}{2 \ell +1} \sum_{m=-\ell }^{m=\ell }
\langle |a_{\ell m}|^{2} \rangle \ . 
\label{cls}
\en
We begin with the contribution --to these coefficients-- of the third
term of the right hand side of Eq. (\ref{efun}), which   
corresponds to the ISW effect. 
An appropriate formula giving this contribution to the
$C_{\ell }$'s has been derived. The most useful feature of this
formula is that, given two $k$--values and two redshifts, it
allows us to obtain a good measure of the ISW 
effect produced by 
the density perturbation located between the chosen redshifts
and having scales between the chosen ones. A few words about the derivation 
--similar to the usual derivation of the $C_{\ell}$ coefficients
of the Sachs-Wolfe effect--
and characteristics of 
this formula are worthwhile.

Since the angular brackets in Eq. (\ref{cls}) stand for a mean 
performed from many realizations of the microwave sky, quantity 
$|a_{\ell m}|^{2}$ is first computed for an observer having comoving 
coordinates
$x^{i}_{_{P}}$ in a reference system attached to the Local Group 
(origin of spatial coordinates) and,
then, an average over position $x^{i}_{_{P}}$ is done to 
get the $C_{\ell} $ quantities.

The equations of the 
null geodesics passing by the origin of 
spatial coordinates are
\be
x^{i}= \lambda_{q} (a) e^{i}  \ . 
\en
Furthermore, in flat cases (as model I), the null geodecis passing by
point $x^{i}_{_{P}}$ are: 
\be
x^{i}=x^{i}_{_{P}} + \lambda_{q} (a) e^{i} \ ,
\label{l7}
\en
while in model II, this last equation is also valid when point P 
is well inside a sphere centred 
at the Local Group and having the size of the curvature scale.
This is because --in such a case-- the spatial curvature 
can be neglected.
       
Using the third term of the right hand side of Eq. (\ref{efun}),
and Eqs. (\ref{l1}),                     
(\ref{l4}) and (\ref{l7}), some Fourier expansions
lead to the following 
relation
\be
\frac {\Delta T}{T} (\vec {x}_{_{P}} , \vec {n})=
\frac {2B_{q}}{(2 \pi)^{3/2}}
\int d^{3} k_{c} e^{-i \vec {k}_{c} \vec {x}_{_{P}}} \frac 
{\delta_{\vec {k}_{c}}}
{k_{c}^{2}}
\int_{_{P}}^{^{Q}}  e^{-i \lambda_{q} (a)  \vec {k}_{c} \vec {n}}
\frac {d} {da} \left[ \frac {D_{1q}(a)}{a} \right] da  \ , 
\label{Four}   
\en
where the components of $\vec {x}_{_{P}}$ are $x^{i}_{_{P}}$, the
observer located at P estimates $\frac {\Delta T}{T}$ in the
direction $\vec {n}$, and point Q is the intersection between
the last scattering surface of P and the null geodesic 
determined by $\vec {n}$.

Finally, from Eq. (\ref{Four}) plus the usual expansions in
spherical harmonics and, after performing the average in $x^{i}_{_{P}}$, 
the following
angular power spectrum arises:
\be
C^{^{I}}_{\ell q }= \Gamma^{^{I}}_{q } 
\int \frac {P(k)} {k^{2}} \xi^{2}_{\ell q } (k) dk  \ , 
\label{isw1}
\en
where
\be
\xi_{\ell q } (k) = \int_{a_{0}}^{a_{_{L}}} 
j_{\ell } [\lambda_{q} (a) k a_{0}]  
\frac {d} {da} \left[ \frac {D_{1q}(a)}{a} \right] da  \ .
\label{isw2}
\en
Function $P(k)= \langle |\delta_{k}|^{2} \rangle$ is the 
power spectrum 
of the energy density fluctuations, 
$j_{\ell}$ is the spherical Bessel function of order $\ell$,
and coefficients 
$\Gamma^{^{I}}_{q }$ are given in Appendix A for models I and II
(all the $\Gamma$ coefficients appearing below are also
listed in the same appendix). 

A similar computation leads to the $C_{\ell}$ coefficients
corresponding to the first and second terms of Eq. (\ref{efun}), which 
are usually referred to as Sachs-Wolfe and Doppler terms. 
In the Sachs-Wolfe case, these 
coefficients can be written as follows:
\be
C^{^{S}}_{\ell q}= \Gamma^{^{S}}_{q}
\int \frac {P(k)} {k^{2}} j_{\ell }^{2} 
[\lambda_{q} (a) k a_{0}] dk
\label{sw1}
\en
and, the coefficients of the Doppler term are
\be
C^{^{D}}_{\ell q}= \Gamma^{^{D}}_{q}   
\int P(k)  j^{\prime 2}_{\ell }          
[\lambda_{q} (a) k a_{0}] dk   \ ,               
\label{d1}
\en
where $j^{\prime}_{\ell } (x) = (d/dx) j_{\ell } (x)$.

Eqs. (\ref{isw1}) -- (\ref{d1}) are written in a form which 
is adequate to perform our numerical estimates. 
We define the functions 
$\mu^{^{I}}_{\ell q}(k)  =    \Gamma^{^{I}}_{q } 
k^{-2} P(k) \xi^{2}_{\ell q } (k)$,
$\mu^{^{S}}_{\ell q}(k)  =    \Gamma^{^{S}}_{q } 
k^{-2} P(k) j_{\ell }^{2}   
[\lambda_{q} (a) k a_{0}]$, and 
$\mu^{^{D}}_{\ell q}(k)  =    \Gamma^{^{D}}_{q } 
P(k) j^{\prime 2}_{\ell }   
[\lambda_{q} (a) k a_{0}]$, whose integrals in the variable $k$
give $C^{^{I}}_{\ell q }$, $C^{^{S}}_{\ell q}$, and 
$C^{^{D}}_{\ell q}$, respectively. These definitions will be
useful below.

If the three terms of the right hand side of Eq. (\ref{efun}) 
are simultaneously taken into account in order to get 
$\langle |a_{\ell m}|^{2} \rangle$,
the resulting $C_{\ell }$ quantities include three crossed
contributions mixing the ISW, SW, and Doppler effects. 
We have not found fully convincing arguments to neglect
these contributions in all the cases and, consequently, they have been 
systematically
estimated using the following formulae:
\be
C^{^{SD}}_{\ell q}= -2
\int \left[ \mu^{^{S}}_{\ell q}(k) 
\mu^{^{D}}_{\ell q}(k) \right]^{1/2} dk   \ ,               
\label{cros1}                                                     
\en                                                              
\be
C^{^{SI}}_{\ell q}= 2
\int \left[ \mu^{^{S}}_{\ell q}(k) 
\mu^{^{I}}_{\ell q}(k) \right]^{1/2} dk   \ , 
\label{cros2}                               
\en
\be
C^{^{DI}}_{\ell q}=-2 
\int \left[ \mu^{^{D}}_{\ell q}(k) 
\mu^{^{I}}_{\ell q}(k) \right]^{1/2} dk   \ . 
\label{cros3}
\en

Since the late ISW effect is produced by 
inhomogeneities evolving after decoupling, quantities
$C^{^{I}}_{\ell q }$ can be estimated using the
above {\em pressureless} approach for the sources;
however, the Sachs-Wolfe and 
Doppler effects are produced by other inhomogeneities, which evolve 
in the recombination-decoupling period and, consequently, 
a certain radiation
pressure is acting on the subdominant baryonic component.
Taking into account that
the importance of pressure effects increases as 
$\ell $ does, we only apply our pressureles approach to calculate
the Sachs-Wolfe, Doppler and crossed coefficients
in the case $\ell = 10$. This calculation is performed 
with the essential aim of obtaining a rough estimate
of the unknown crossed terms and, for this purpose, our 
approach suffices.

In the $\ell$ interval [10,40], we can only expect 
significant contributions 
to the $C_{\ell}$ coefficients coming
from: (1) a possible background of
primordial gravitational waves (this contribution would be almost 
independent on $\ell$ in the interval under consideration and
it is not studied here), (2)
each of the three effects 
considered above and, (3) some crossed terms. 
Other effects as Sunyaev--Zel'dovich, lens anisotropy, 
nonlinear gravitational
effects et cettera are not expected to be relevant 
for these angular scales, but for much smaller ones.

The sources of the term 
$C^{^{I}}_{\ell q} $ 
have been identified in both scale and position using 
the following definitions:
\be
D^{^{I}}_{\ell q}(Z_{min}, Z_{max})=\Gamma^{^{I}}_{q}
\int \frac {P(k)} {k^{2}} \zeta^{2}_{\ell q} 
(k, Z_{min}, Z_{max}) dk  \ , 
\label{isw1b}
\en
where
\be
\zeta_{\ell q} (k, Z_{min}, Z_{max}) = a_{0}^{-1} \int_{Z_{min}}^{Z_{max}}   
j_{\ell } \left[ \lambda_{q} 
\left( \frac {a_{0}}{1+Z} \right) k a_{0} \right]  
\frac {d} {dZ} \left[ D_{1 q} \left( \frac {a_{0}}{1+Z} \right) (1+ Z)
\right] dZ   \ .
\label{isw2b}
\en                      
For $Z_{min}=0$ and $Z_{max} = Z_{_{L}}$, where  
$Z_{_{L}}$ is the redshift 
of the last scattering surface, functions 
$\zeta_{\ell q} (k, Z_{min}, Z_{max})$
and $D^{^{I}}_{\ell q}(Z_{min}, Z_{max})$ are identical to 
$\xi_{\ell q} (k)$ and $C^{^{I}}_{\ell q}$,
respectively.
Quantity $D^{^{I}}_{\ell q}(Z_{min}, Z_{max})$ can
be considered as a measure of
the contribution --to the ISW effect-- of the inhomogeneities 
lying between redshifts $Z_{min}$ and $Z_{max}$; nevertheless,
it is worthwhile to emphasize that the right hand side of Eq. (\ref{isw1b})
involves the function $\zeta^{2}_{\ell q} (k, Z_{min}, Z_{max})$,
which implies that the
quantities $C^{^{I}}_{\ell q}=D^{^{I}}_{\ell q}(0, Z_{_{L}})$ 
are not the linear superposition 
of quantities of the form $D^{^{I}}_{\ell q}(Z_{min}, Z_{max})$, 
even if these
quantities are calculated in 
disjoint redshift intervals covering the total interval
($0,Z_{_{L}}$). 
From Eq. (\ref{isw1b}) it follows that 
the contribution of each scale to $D^{^{I}}_{\ell q}(Z_{min}, Z_{max})$
is measured by the function
$\nu^{^{I}}_{\ell q}(k, Z_{min}, Z_{max}) = 
\Gamma^{^{I}}_{\ell q} P(k) \zeta^{2}_{\ell q} 
(k, Z_{min}, Z_{max}) /k^{2}$. This function measures the contribution 
of the scale $k$ --for
the inhomogeneities placed between redshift $Z_{min}$ and $Z_{max}$--  
to the late ISW effect .
For $Z_{min}=0$ and $Z_{max} = Z_{_{L}}$, function 
$\nu^{^{I}}_{\ell q}(k, Z_{min}, Z_{max})$ 
is identical to function $\mu^{^{I}}_{\ell q}(k)$ and it
weights the 
contribution of each scale --whatever the inhomogeneity location may be--    
to the ISW angular power spectrum.  

Our calculations require a value of $Z_{_{L}}$.
Since the Sachs-Wolfe and Doppler effects are produced by inhomogeneities
located very near the last scattering surface, estimates of
$C^{^{S}}_{\ell q}$ and $C^{^{D}}_{\ell q}$ based on Eqs. (\ref{d1})
and (\ref{sw1})
are sensitive to the value of $Z_{_{L}}$; however, the late
ISW effect is produced by inhomogeneities 
located far from this surface (see Section 3) and,
consequently, it is
almost independent on the assumed value
of $Z_{_{L}}$. In order to do the best estimate of the
Doppler and SW effects for $\ell = 10$ --allowed by our formalism--          
we have taken $Z_{_{L}}=1140$, which is the redshift corresponding
to $\Omega_{b}=0.05$ and $\Omega_{d}=0.25$ according to the formula
$Z_{_{L}} \simeq 1100 (\Omega_{m} / \Omega_{b})^{0.018}$
(see Kolb \& Turner 1994). Fortunately, we are focusing our 
attention on the late ISW effect, which is almost independent on
the choice of $Z_{_{L}}$.

\section{RESULTS}

Assuming the normalization $\sigma_{8}=1$, quantities
$C^{^{I}}_{\ell q}$, $C^{^{S}}_{10 q}$,
$C^{^{D}}_{10 q}$, $C^{^{SD}}_{10 q}$, $C^{^{SI}}_{10 q}$, 
$C^{^{DI}}_{10 q}$, and $C_{10 q}$ have been computed in models I (q=I)
and II (q=II). For this first normalization, 
Quantity
$[\ell (\ell + 1) C^{^{I}}_{\ell q} / 2 \pi ]^{1/2}$ 
is shown in the left panel of
Fig. 1 (for models I and II). The entries 1 and 2 of Table 1
gives  
$[110 C_{10} / 2 \pi ]^{1/2}$ for all the 
$C_{10}$ quantities. In this Table we see that:
for $\ell = 10$ and model I, the ISW effect is smaller than 
the Doppler and SW ones, while for $\ell = 10$ and model II,
the ISW and the SW effects are similar.
A different normalization facilitates some comparisons 
of the anisotropies appearing in 
models I and II. We have observed that most 
theoretical predictions   
based on COBE normalization give
$[ \ell (\ell + 1) C_{\ell}]^{1/2} \sim 28 \ \mu K$ for $\ell = 10$,
with a small dispersion around 28 $\mu K$. This is true 
for a wide range of variation of the cosmological
parameters: $\Omega_{0}$, $\Omega_{b}$ et cettera.
This condition is also 
compatible with all the  
observational evidences (FIRS, TENERIFE).
By these reasons,  the  ISW
effects corresponding to models I and II with the normalization 
$[ \ell (\ell + 1) C_{\ell}]^{1/2} = 28 \ \mu K$ for $\ell = 10$ 
are represented in the right panel of Fig. 1, where 
we see that: (i) the ISW
effect corresponding to model I (with cosmological
constant) is much smaller
than that of the model II (very open universe), and (ii) 
the ISW effect of model I is small but it is not negligible.
Entries 3 and 4 
of Table 1 correspond to 
the second normalization,  
for which,  the bias parameter
of model I (II) appears to be 0.93 (1.98). This means that
the currently preferred model (with cosmological
constant) leads to a very natural compatibility 
between the
CMB observational data and the value $\sigma_{8} \sim 1$ 
extracted from the  analysis of galaxy surveys.

In scenario I, the greatest $\ell = 10$  crossed term is 
the SW--Doppler one ($C^{^{SD}}_{10 I}$), 
which is shown in the entries 1 and 3 of Table 1. The
remaining
crossed terms (SW-ISW and Doppler-ISW)
are not given because they have appeared 
to be negligible. 
In entries 3 and 4 of Table 1, we give the SW-ISW crossed
term ($C^{^{SI}}_{10 II}$) of the scenario II, which is not negligible;
however, the terms Doppler--SW and Doppler--ISW
can be neglected. 
As it follows from Eqs. 
(\ref{cros1}) -- (\ref{cros3}),
any crossed term is proportional to 
an integral (in the variable $k$), and the function 
to be integrated can be written as the product of two $k$ functions
corresponding to the mixed effects. For example,
in the SW--Doppler (SW--ISW) case, we must integrate the
product $\left[ \mu^{^{S}}_{\ell q}(k) \right]^{1/2}
\left[ \mu^{^{D}}_{\ell q}(k) \right]^{1/2}$
($\left[ \mu^{^{S}}_{\ell q}(k) \right]^{1/2}
\left[ \mu^{^{I}}_{\ell q}(k) \right]^{1/2}$).
In Fig. 2, we display the functions to be multiplied
to get the SW-Doppler crossed term of model I 
(left panel) and the SW-ISW term of model II (right panel).
We have assumed 
$\ell = 10$ in both models,   
evidently, these 
crossed terms are not negligible
as a result of the existence of a wide enough $k$ interval where 
the positive functions to be multiplied take on large enough 
values simultaneously. 
              
Where are located the inhomogeneities producing the ISW effect?
This question can be answered using Eqs.
(\ref{isw1b}) and (\ref{isw2b}) to calculate 
$D^{^{I}}_{\ell q}(0, Z_{max})$ for appropriate values of
$Z_{max}$. Results are shown in 
Fig. 3, where $D^{^{I}}_{\ell q}(0, Z_{max})$ is represented as a function
of $Z_{max}$ in models I (top) and II (bottom). 
The points of the horizontal straight lines
of Fig. 3 have the ordinate 
$C^{^{I}}_{\ell q}=D^{^{I}}_{\ell I}(0, 1140)$ and, 
consequently, the curves $D^{^{I}}_{\ell q}(0, Z_{max})$ 
must tend to the horizontal lines as 
$Z_{max}$ tends to $1140$.
In the top panel, we see that 
$D^{^{I}}_{\ell I}(0, Z_{max})$ approaches the horizontal
lines very quickly.
Quantity $D^{^{I}}_{\ell I}(0,2)$
is very similar to quantity
$C^{^{I}}_{\ell q}=D^{^{I}}_{\ell I}(0, 1140)$,
which means that the most part of the late ISW is 
produced by inhomogeneities 
located at very low redshift ($Z \leq 2$). 
This is because the cosmological constant is known to be
significant only at very low redshifts; before,
the universe can be considered as a
flat one with a negligible cosmological constant, and
no ISW effect is expected in this situation. 
In model II (bottom panel), we see that
$D^{^{I}}_{\ell II}(0, Z_{max})$ approaches the
corresponding horizontal line more slowly than in model I.
The most important part of the ISW effect is produced 
by inhomogeneities located at redshift $Z < 10$, 
but inhomogeneities at $Z > 10$ also produce a small but
appreciable effect.

Now, let us look for the spatial scales contributing 
significantly to the ISW effect in the $\ell $ interval [10,40].
As stated before, the contribution of the scale $k$ to the 
ISW effect --for arbitrary location of the inhomogeneities--
is weighted by the function 
$\mu^{_{I}}_{\ell q}(k)$. In Fig. 4, this function is represented 
with solid lines in
cases I (top) and II (bottom). Left (right)
panels correspond to $\ell = 10$ ($\ell = 40$).                  
Taking into account that, for $h = 0.65$, 
the spatial size (diameter in the spherical case)
of the structures 
associated to the wavenumber $k$  
is $\frac {2}{k} h^{-1} \ Mpc$,
Fig. 4 can be easily interpreted. In each panel, the solid
lines show a $k$ value for which 
the function $\mu^{_{I}}_{\ell q}(k)$ reaches a maximum.
The spatial scale corresponding to this 
$k$ value is hereafter denoted $D^{*}$.
It is the most significant scale for ISW anisotropy 
production.
Solid lines also show the existence of a minimum
$k$ value where $\mu^{_{I}}_{\ell q}(k)$ starts to
increase from negligible values. The spatial scale
associated to the minimum will be denoted $D_{max}$.
Scales larger than this maximum one do not contribute
to the ISW effect significantly.
The scales $D^{*}$ and $D_{max}$ corresponding to
the four solid lines of Fig. 4 are given in Table 2.
The meaning of these scales is discussed in next section.

Finally, among all the inhomogeneities located between 
redshifts $Z_{min}=0$ and $Z_{max}$, which of them 
are contributing to the
ISW effect? Which are the spatial scales of these 
inhomogeneities?
In order to answer this question we have put
$Z_{min}=0$ and 
various values of $Z_{max}$ into Eq. (\ref{isw2b}); thus, 
we have found various  functions 
$\nu^{^{I}}_{\ell q}(k,0,Z_{max})$ of the variable $k$. 
Only two of these functions are displayed in each panel of Fig. 4.   
The function corresponding to $Z_{max} = 1140$
is identical to
$\mu^{^{I}}_{\ell q}(k)$ 
and it is drawn 
with solid lines, while
the dotted lines correspond to $Z_{max}=0.5$ in model
I (top panels) and to $Z_{max}=2$ in model II (bottom panels).
In Fig. 3, we see that for these $Z_{max}$ values, a significant
part of the ISW effect is produced by inhomogeneities 
located at $Z<Z_{max}$. The dotted lines of 
Fig. 4 lead --as the solid lines-- to new values of 
$D^{*}$ and $D_{max}$ 
which are presented in Table 2 and interpreted below.
These scales correspond to inhomogeneities located
at low redshifts smaller than the chosen values of $Z_{max}$
             
\section {DISCUSSION AND CONCLUSIONS} 

In this paper, we have accurately
estimated the scales and locations of the inhomogeneities 
contributing to the late ISW effect. The chosen formalism
has facilitated our analysis. Now, let us focus our 
attention on the meaning of the resulting scales
(see Table 2).
They are not the scales 
of the inhomogeneities (density contrasts)
producing the effect. According to 
Eq. (\ref{l1} ), the ISW effect is produced by the scales 
contributing significantly to the partial time derivative of
the peculiar gravitational 
potential and, 
in the linear regime under consideration,
these scales are identical to those of the 
potential itself (see Eq. (\ref{l4})).

The significant spatial scales of an overdensity and those of its peculiar 
gravitational potential are different,               
this is proved by the relation $\phi_{\vec{k}} \propto 
\delta_{\vec{k}} / k^{2}$
between the Fourier transforms of the density contrast $\delta_{\vec{k}}$
and the peculiar potential $\phi_{\vec{k}}$. The factor $1/k^{2}$
implies that the regions where the potential is significant 
are more extended 
than those where the density contrast is not negligible.
What is the size of the regions where the potential is
contributing to the late ISW effect?
In order to answer this question let us consider
a spherically symmetric 
overdensity. In such a case, Eq. (\ref{lapla}) leads to the relation
\be
\frac {\partial \phi}{\partial r} \propto \frac {M(r)}{r^{2}} \ , 
\label{fir}
\en
where $M(r)$ is the total mass inside a sphere of radius $r$.
This relation
allows us to get various important features of the peculiar 
gravitational potential generated by a compensated overdensity. In fact,
according to the cosmological principle, any overdensity must be compensated 
at some distance, $r_{c}$, from its center, at which the total mass excess
$M(r_{c})$ vanishes. This excess also vanishes for $r > r_{c}$. Then, 
according to Eq. (\ref{fir}), the derivative 
$\frac {\partial \phi}{\partial r} (r_{c})$ vanishes for
$r > r_{c}$ and, consequently, the potential reaches a minimum 
constant value, which should be zero to achieve good boundary
conditions at infinity. We see that the potential of a 
compensated structure tends to zero as $r$ tends to $r_{c}$; 
hence, all the shells forming a certain structure 
--up to compensation radius-- would contribute to the 
ISW effect, although this contribution would be small 
for r values close to $r_{c}$; 
hence, the late ISW effect produced by a given structure
depends on the way in which it is compensated; namely,
it depends on the size $\sim 2r_{c}$ of the region where 
the potential is contributing to the ISW effect. This 
region is hereafter referred as to the pot-region associated to
the inhomogeneity.
Since we are considering linear scales where the peculiar velocities
are proportional to the gradients of the peculiar gravitational
potential, these velocities also vanish for $r>r_{c}$ and, consequently,
the pot-region contributing to the ISW effect is that where the
peculiar velocities (potential gradients) are significant. 
A given overdensity would
contribute to 
a certain $C_{\ell}$ coefficient, 
if the angular scale subtended by its pot-region (not the angular scale
subtended by itself)
is appropriate. 

The compensation of cosmological objects is a
statistical phenomenon and, consequently, structures
of the same type
(for example various GAL structures)
could be compensated at different distances from their cores;
therefore, although
the compensation radius of the Great Attractor has been estimated 
to be $r_{c}^{^{GA}} \sim 100h^{-1} \ Mpc$
(study of the velocity field around the GA), other GAL
structures could compensate 
at other distances, perhaps at distances of a few times $r_{c}^{^{GA}}$.
Voids and Abell clusters would compensate at distances of a
few tens of Mpc from the central region. 
Taking into account these considerations we
are going to interpret the results summarized in Table 2.

For model I and $\ell = 40$ ($\ell = 10$), pot-regions with radius 
larger than  $250 h^{-1} \ Mpc$ ($1000 h^{-1} \ Mpc$)
are not contributing
to the late ISW effect. 
Those having radius larger than $50 h^{-1} \ Mpc$ ($200 h^{-1} \ Mpc$)
and located at $Z<0.5$ do not contribute either.
The maximum effect is produced by pot-regions with 
radius close to $100 h^{-1} \ Mpc$ ($300 h^{-1} \ Mpc$)
and located between 
redshifts $0.5$ and $2.$ and, finally, the maximum 
effect produced at $Z<0.5$ is due to pot-regions
having about $40h^{-1} \ Mpc$ ($140 h^{-1} \ Mpc$) radius.
This means that, in model I, GAL structures 
with $r_{c} \sim 100 h^{-1} \ Mpc$ produce
the maximum contribution to $C_{40}$ (the smallest of the 
$C^{^{I}}_{\ell I}$ coefficients, see the top panel of Fig. 1).
GAL objects with sizes $r_{c} \sim 140 h^{-1}$ would
contribute to $C_{10}$ when located at $Z<0.5$ and
GAL objects with pot-regions of a few times $100 h^{-1} \ Mpc$
would contribute to all the
$C^{^{I}}_{\ell I}$ coefficients from $\ell=10$ to $\ell=40$.
Pot-regions with radius of $\sim 40 h^{-1} \ Mpc$ and
located at $Z<0.5$ contribute to $C_{40}$.

In model II, a similar study has been developed. 
For $\ell = 40$ ($\ell = 10$), the following 
conclusions can be obtained: 
(i) for arbitrary locations, 
the most large compensation radius contributing to the late ISW is
$650 h^{-1} \ Mpc$ ($2500 h^{-1} \ Mpc$),    
(ii) for structures located at $Z<2$, 
the most large radius is
$125 h^{-1} \ Mpc$ ($500 h^{-1} \ Mpc$),  
(iii) for arbitrary locations, 
the most large contributions to the late ISW come from 
compensation radius of
$300 h^{-1} \ Mpc$ ($1000 h^{-1} \ Mpc$), and
(iv) for structures located at $Z<2$, 
the most large contributions correspond to
radius of $\sim 110h^{-1} \ Mpc$ ($\sim 350h^{-1} \ Mpc$).
In model II, the scales are larger than those of model I.
Scales of a few times $\sim 10h^{-1} \ Mpc$ do not play any role.
GAL structures 
with $r_{c} \sim 100 h^{-1} \ Mpc$ only contribute to $C_{40}$
if located at $Z<2$. 
GAL objects compensated at radius around $300 h^{-1} \ Mpc$
would play an important role in generating 
the $C^{^{I}}_{\ell I}$ coefficients, in particular,
for $\ell = 40$. This conclusion is in agreement with 
previous claims about the possible relevance of 
GAL structures in generating the late ISW effect in open 
universes (Arnau, Fullana \& S\'aez, 1994;
S\'aez, Arnau \& Fullana, 1995).
The GAL objects simulated in those studies (based on TB)
undergo
effective compensations at distances of 
a few hundred of Megaparsec from their cores and, 
in agreement with the results of this paper (Table 2), 
this type of structures would be contributing significantly to 
the late ISW effect.

It is worthwhile to emphasize that
we have discussed the contribution to the late ISW effect of
great cosmological structures
(which produce peculiar velocities up to distances of
tens or hundreds of Mpc).
A linear approach suffices to estimate 
the potential (also the peculiar velocities) produced
by these structures (GAL objects, voids et cettera).
We have never considered the Rees-Sciama effect produced by
strongly nonlinear substructures lying inside the 
Great Attractor and other extended inhomogeneities.
Such an effect would produce CMB anisotropy on smaller
angular scales and its estimate would require other 
nonlinear approaches.

\vspace {1 cm}
\noindent
{\large{\bf APPENDIX A}}\\
\
Some quantities used in this paper have the 
subscript "$q$". Here, the explicit form of these quantites
is given for 
$q=I$ (model I of Section 1) and $q=II$ (model II).
We summarize the information as follows:

MODEL I ($q=I$)

The growing mode of the scalar energy density fluctuations
is
\be
D_{1I}(a) = \frac {1}{x} \left[ \frac {2}{x} + x^{2} \right]^{1/2}
\int_{0}^{x} \left[ \frac {2}{y} + y^{2} \right]^{-3/2} dy  \ , 
\label{ll2}       
\en
where
\be
x = \left[ \frac {2 \Omega_{\lambda}}{\Omega_{m}} \right]^{1/3}
(1+Z)^{-1} \ . 
\en

The constant $B_{q}$ is   
\be
B_{I} =  - \frac {3} {2} \frac {\Omega_{m} H_{0}^{2}} {D_{1}(0)} \ . 
\label{ll5}        
\en

The function $\lambda_{q} (a)$ can be 
written as follows:
\be
\lambda_{I} (a)= \kappa (a) H_{0}^{-1} \ ,               
\en     
where
\be
\kappa(a) = \int_{a}^{1} \frac {db} {(\Omega_{m0}b+
\Omega_{\lambda} b^{4})^{1/2}} \ .
\en

Now, we give the coefficients $\Gamma^{^{I}}_{I }$,
$\Gamma^{^{S}}_{I}$,
$\Gamma^{^{D}}_{I}$,
$\Gamma^{^{SD}}_{I}$,
$\Gamma^{^{SI}}_{I}$, and
$\Gamma^{^{DI}}_{I}$ defined in Section 2.

\be
\Gamma^{^{I}}_{I } = \frac {18H_{0}^{4}} {\pi}
\left[ \frac {\Omega_{m0}} {D_{1}(0)} \right]^{2} \ , 
\en

\be
\Gamma^{^{S}}_{I}= \frac {H_{0}^{4}} {2 \pi} 
\left[ \frac {D_{1}(L) \Omega_{m0} (1+Z_{_{L}})} {D_{1}(0)} 
\right]^{2} \ ,  
\en

\be
\Gamma^{^{D}}_{I} = \frac {2 H_{0}^{2}} {\pi}
\left[ \frac {D_{1}(L)} {D_{1}(0)} \right]^{2}  
\Omega_{m0} (1+Z_{_{L}})  \ . 
\en

In order to derive these formulae, the cosmological constant
has been assumed to be negligible at $Z_{_{L}}$ (see
comments of Section 3) and, consequently,
as $\Omega_{m0}$ tends to unity, quantities 
$\Gamma^{^{S}}_{I}$ and $\Gamma^{^{D}}_{I}$ tend to
the right values corresponding to a flat universe
without cosmological constant
($H_{0}^{4} / 2 \pi$ and $2H_{0}^{2}/ \pi (1+Z_{_{L}})$).

MODEL II ($q=II$)

The same quantities as in model I are now listed: 
\be
D_{1II}(a)=1 + \frac {3}{\zeta } + \frac {3(1+\zeta )^{1/2}}{\zeta^{3/2}}
ln [ (1+\zeta)^{1/2} -\zeta^{1/2} ]  \ , 
\label{l2}       
\en
where
\be
\zeta = \frac {H_{0} (1 - \Omega_{0})^{3/2}}{\Omega_{0}} a \ .
\label{l3}       
\en

\be
B_{II} = - \frac {3}{2} \Omega_{0} \left[ H_{0} D_{1} (a_{0}) 
(1- \Omega_{0})^{3/2} \right]^{-1} \ .
\label{l5}        
\en

\be
\lambda_{II} (a)= 2 \tanh (y/2)     \ ,               
\en     
where 
\be
y=\cosh^{-1} \left[ \frac {2- \Omega_{0}}{ \Omega_{0} } \right] -
\cosh^{-1} \left[\frac {2 (1- \Omega_{0})^{3/2} H_{0}}
{\Omega_{0}} a +1) \right]     \ .
\en

\be
\Gamma^{^{I}}_{II } =
\frac {18 \Omega_{0}^{2} 
H_{0}^{2}} {\pi (1-\Omega_{0})
D_{1}^{2} (a_{0})}  \ , 
\en

\be
\Gamma^{^{S}}_{II}=
\frac {H_{0}^{4}}{2 \pi} 
\Omega_{0}^{2} (1+Z_{_{L}})^{2}
\frac {D_{1}^{2}(a_{_{L}})} {D_{1}^{2}(a_{0})}   \ , 
\label{sw11}
\en

\be
\Gamma^{^{D}}_{II} =
\frac {2 \Omega_{0}}
{\pi (1- \Omega_{0}) D_{1}^{2}(0)  
(1 + Z_{_{L}})} \left( \frac {dD_{1}}{da} \right)^{2}_{_{L}} \ . 
\label{d11}
\en

As $\Omega_{0}$ tends to unity,
function $D_{1}(a)$ tends to 
the growing mode of a flat background, which is
proportional to $a$. Taking into account this fact
and Eqs. (\ref{sw11}) and 
(\ref{d11}), one easily concludes that --as the universe
approaches a flat one-- quantities
$\Gamma^{^{S}}_{II}$ and $\Gamma^{^{D}}_{II}$ tend to
$H_{0}^{4} / 2 \pi$ and $2H_{0}^{2}/ \pi (1+Z_{_{L}})$, 
respectively. These limit values coincide with the
well known values of $\Gamma^{^{S}}_{II}$ and
$\Gamma^{^{D}}_{II}$ corresponding to the 
flat background without cosmological constant.

\newpage

\noindent
{\large{\bf REFERENCES}}\\
\\ 
Arnau J.V., Fullana M.J., S\'aez D., 1994, 
MNRAS, 268, L17\\  
Hu, W., \& Sugiyama, N., 1994, Phys. Rev. 50D, 627\\
Kamionkowski, M., \& Spergel, D.N., 1994, ApJ, 432, 7\\ 
Knox, L., 1995, Phys. Rev., 52D, 4307\\
Kolb, E.W., \& Turner, M.S., 1994, {\em The Early Universe},
Addisson Wesley\\  
Peebles, P.J.E., 1980, {\em The Large Scale Structure of the Universe},
Princeton University Press\\
Mart\'{\i}nez-Gonz\'alez, E., Sanz, J.L. Silk, J., 1994, ApJ, 436, 1\\
S\'aez D., Arnau J.V., Fullana M.J., 1995, Astro. Lett. and Communications,
32, 75\\   
S\'aez D., Fullana M.J., 1999, Astro. Lett. and Communications, in press\\
Sanz, J.L., Mart\'{\i}nez-Gonz\'alez, Cay\'on, L., Silk, J.L.,
Sugiyama, N., 1996, ApJ, 467, 485\\

\noindent
{\bf ACKNOWLEDGMENTS}. This work has been partially
supported by the Spanish DGES (project PB96-0797). 

\newpage

\begin{center}
{\bf FIGURE CAPTIONS}
\end{center}

\vskip 0.5cm
\noindent

\noindent
{\bf FIG.\ 1.--} Each panel shows the quantity
$[\ell (\ell + 1) C^{I}_{\ell q} / 2 \pi ]^{1/2}$ (in $\mu K$) as a 
function of $\ell$. Left (right) panel corresponds to the normalization 
$\sigma_{8}=1$ 
($[ \ell (\ell + 1) C_{\ell}]^{1/2} = 28 \ \mu K$ for $\ell = 10$). 

\vskip 0.5cm

\noindent
{\bf FIG.\ 2.--} Left panel shows 
the quantity $\mu_{10I}^{1/2} \times 10^{5}$ defined
in the text as a function of $k$ for the SW (solid)
and Doppler (dotted) effects. In the right panel,
quantity $\mu_{10I}^{1/2} \times 10^{4}$  
is plotted vs. $k$ 
for the SW (solid) and ISW (dotted) effects.
These functions must be multiplied to get 
the corresponding crossed contributions to the
CMB angular power spectrum.

\vskip 0.5cm

\noindent
{\bf FIG.\ 3.--} Top panel shows 
the quantity $D^{^{I}}_{\ell I}(0,Z_{max})$ defined
in the text as a function of $Z_{max}$ for
$\ell=10$ (solid, not horizontal) and $\ell=40$
(dotted, not horizontal). 
These curves approach the horizontal lines of the 
same type as $Z_{max}$ increases, crossing them at
$Z_{max}=1140$. Bottom panel has the same structure,
but it shows the quantity
$D^{^{I}}_{\ell II}(0,Z_{max})$.

\vskip 0.5cm  

\noindent
{\bf FIG.\ 4.--} Top left: plot of $\nu^{^{I}}_{10I}(k,0,Z_{max})$ 
(see text) vs. $k$ for $Z_{max}=1140$ (solid) and $Z_{max}=0.5$
(dotted); top right: the same as in top left panel for
the quantity $\nu^{^{I}}_{40I}(k,0,Z_{max})$;
bottom left: plot of $\nu^{^{I}}_{10II}(k,0,Z_{max})$ vs. $k$ 
for $Z_{max}=1140$ (solid) and $Z_{min}=2$ (dotted); 
and bottom right:
the same as in bottom left for $\nu^{^{I}}_{40II}(k,0,Z_{max})$.

\vskip 0.5cm  

\newpage

\begin{table}
\begin{center}
{\bf TABLE 1}\\ 
$[\ell (\ell + 1) C_{\ell} / 2 \pi ]^{1/2}$ (IN $\mu K$) 
FOR  $\ell =10$ \\
\begin{tabular}{cccccccc}\\
\hline
\hline 
MODEL & $\sigma_{8}$ & ISW &SW & Doppler & SW--Doppler & SW--ISW & Total\\
\hline
I & 1.00 & 6.75 & 21.85 & 12.63 & 1.25  & -- & 26.15\\  
II & 1.00 & 38.18 & 38.03 & 12.63 & --  & 12.30 & 55.34\\ 
I & 1.07 & 7.23 & 23.39 & 13.52 & 1.34  & -- & 28.00\\  
II & 0.51 & 19.32 & 19.24 & 6.39 & --  & 6.22 & 28.00\\ 
\hline
\multicolumn{8}{c}{}\\
\end{tabular}
\end{center}
\end{table} 

\newpage

\begin{table}
\begin{center}
{\bf TABLE 2}\\ 
PRESENT SIZES OF THE INHOMOGENEITIES \\
PRODUCING THE LATE ISW EFFECT\\
\begin{tabular}{ccccc}\\
\hline
\hline 
MODEL & $\ell$ & $Z_{max}$ & $D_{max}$ & $D^{*}$\\
      &        &     & $h^{-1} \ Mpc$ & $h^{-1} \ Mpc$\\
\hline
I & 10 & 1140 & 2000  & 600\\  
I & 10 & 0.5 & 400 & 280\\  
I & 40 & 1140 & 500 & 200\\ 
I & 40 & 0.5 & 100 & 85\\     
II & 10 & 1140 & 5000  & 2000\\
II & 10 & 2. & 1000 & 700\\
II & 40 & 1140 & 1300 & 600\\
II & 40 & 2. & 250 & 220\\
\hline
\multicolumn{5}{c}{}\\
\end{tabular}
\end{center}
\end{table} 
  
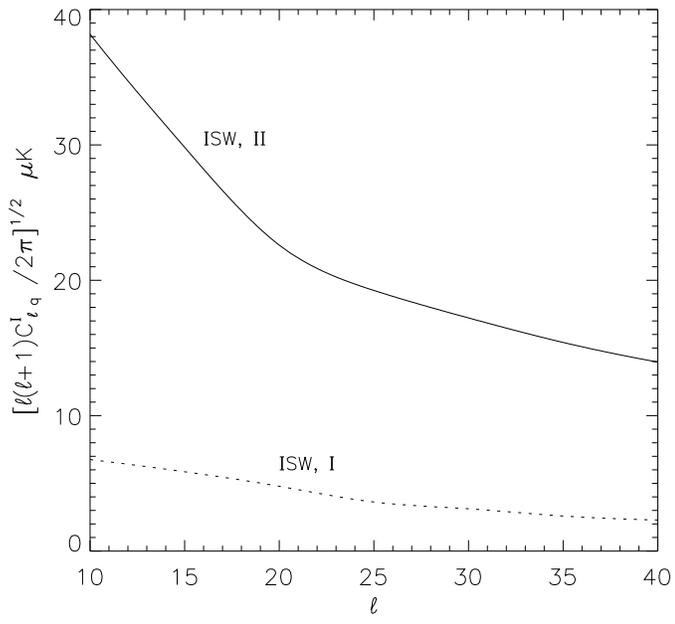
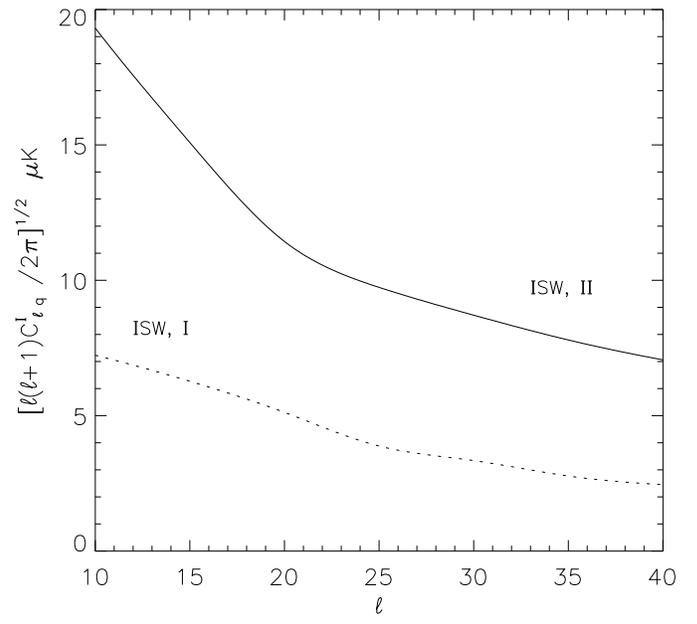
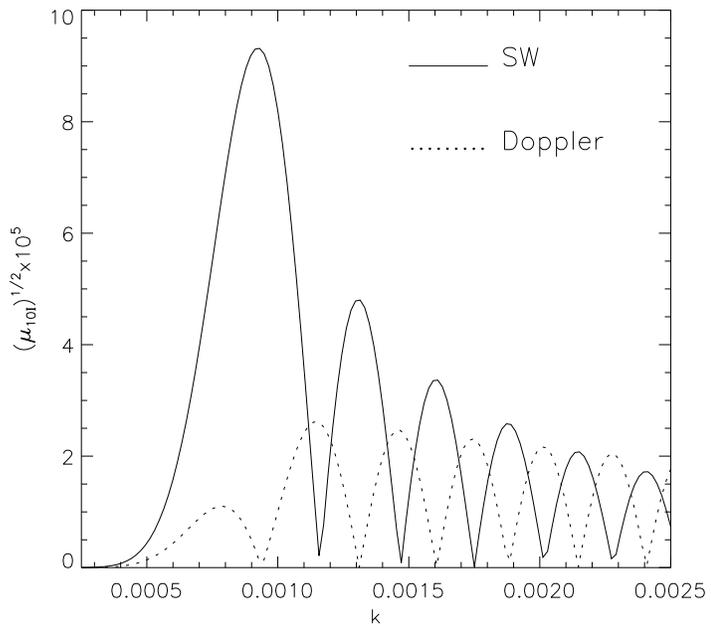
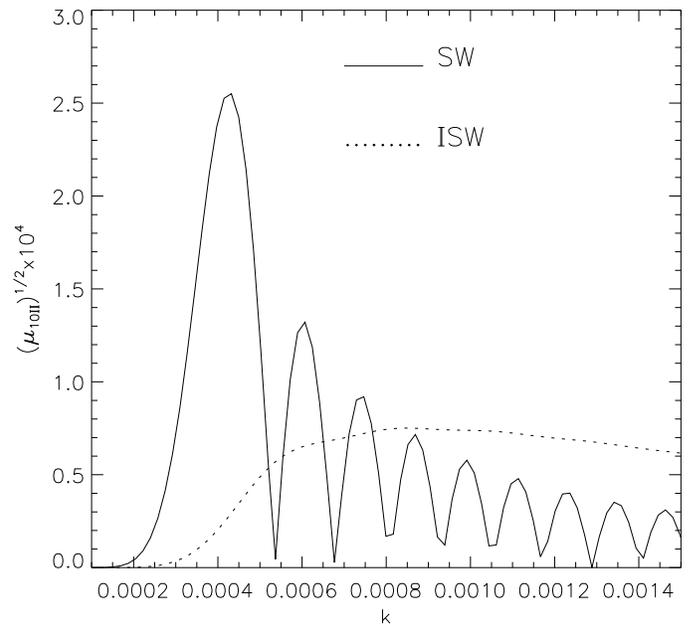
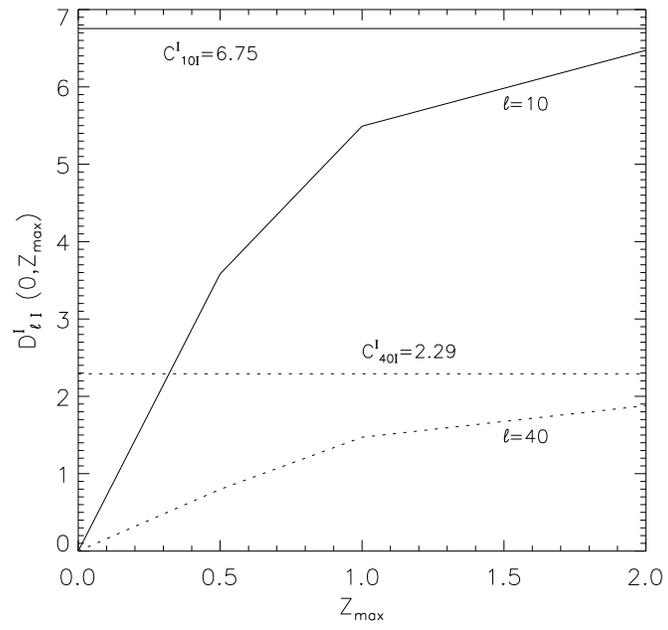
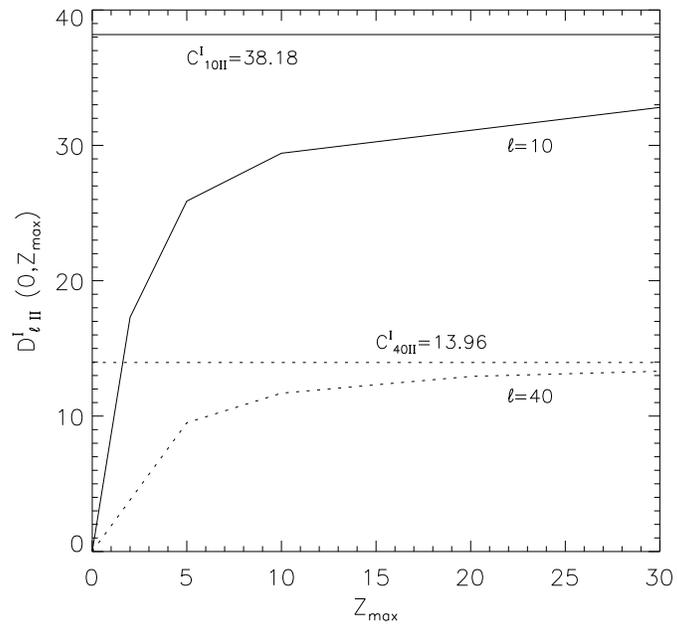
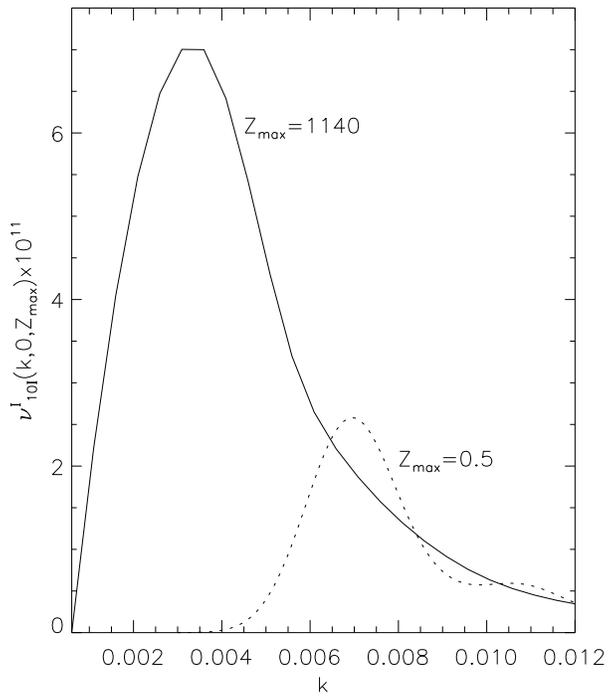
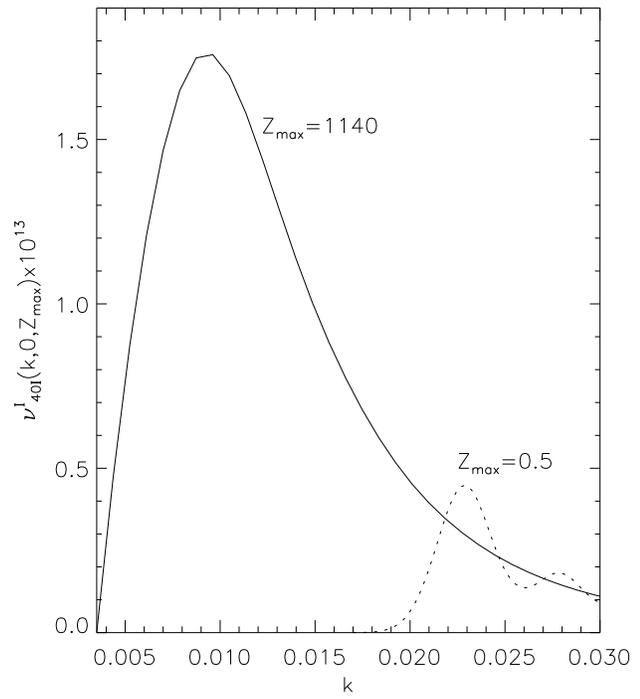
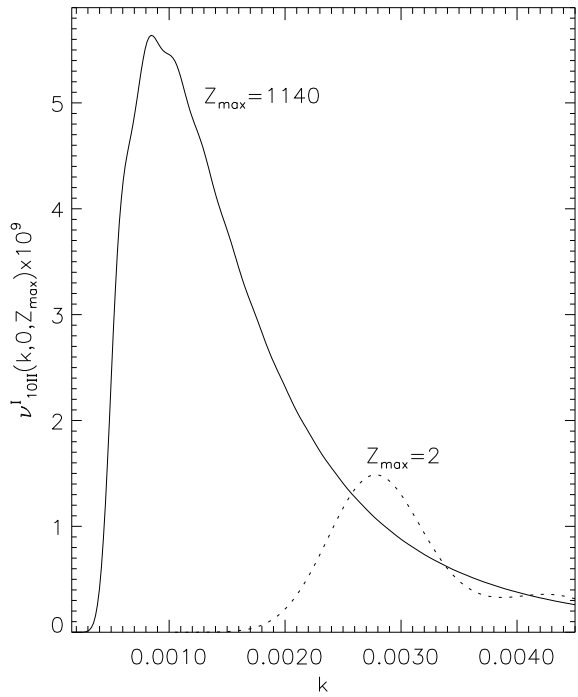
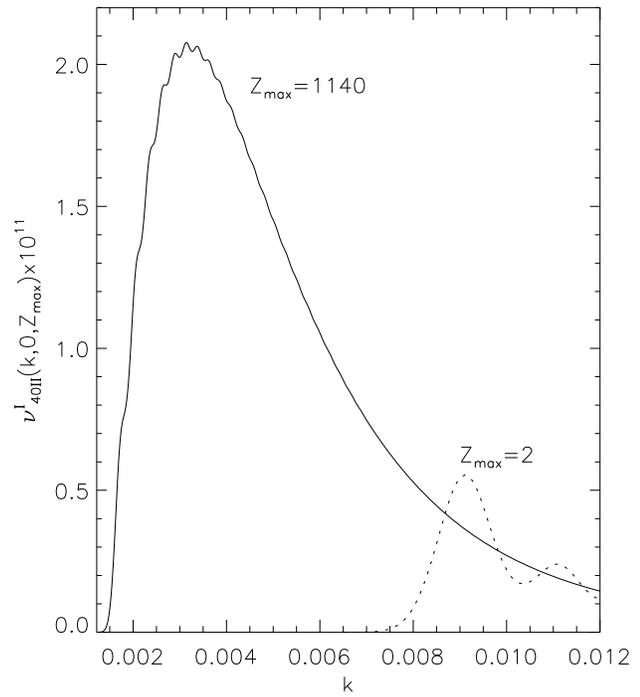
\end{document}